\documentclass[12pt]{article}
\usepackage{amsmath}
\usepackage{amsfonts}
\usepackage{amssymb}
\usepackage{latexsym}
\usepackage{graphicx}
\usepackage[english]{babel}
\input epsf.sty

\newcommand{\be}{\begin{equation}}
\newcommand{\ee}{\end{equation}}
\newcommand{\ba}{\begin{array}}
\newcommand{\ea}{\end{array}}
\newcommand{\bqa}{\begin{eqnarray}}
\newcommand{\eqa}{\end{eqnarray}}

\renewcommand{\d}{\mathrm{d}}
\renewcommand{\H}{\mathcal{H}}

\begin{document}

\title{Revisiting non-Gaussianity of multiple-field inflation from the
field equation} \vspace{3mm}

\author{Shi-Wen Li$^{1}$, Wei Xue$^{1}$\\
{\small $^{1}$ School of Physics, Peking University, Beijing 100871,
China}}

\date{}
\maketitle

\begin{abstract}

In the present paper, we study the non-Gaussianity of multiple-field
inflation model using the method of the field equation. We start
from reviewing the background and the perturbation theory of
multiple-field inflation, and then derive the Klein-Gorden equation
for the perturbations at second order. Afterward, we calculate the
tree-level bispectrum of the fields' perturbations and finally give
the corresponding parameter $f_{NL}$ for the curvature perturbation
$\zeta$ in virtue of the $\delta N$ formalism. We also compare our
result with the one already obtained from the Lagrangian formalism,
and find they are consistent. This work may help us understand
perturbation theory of inflation more deeply.

\end{abstract}

\newpage

\section{Introduction}

It is suggested that our universe has undergone an inflationary
stage in the early time. This scenario helps us understand why our
universe are so flat and isotropic, and also provides a possible
solution to the monopole problem in the hot Big-Bang
cosmology~\cite{Guth81,Linde82,Steinhardt82}. The most efficient
model of inflation is driven by a single scalar field which rolls
down along its potential very slowly. This model generically
predicts a scale-invariant powr spectrum and so is able to explain
the formation of the large scale structure. This expectation has
already been confirmed by the 5-year WMAP data~\cite{wmap} which is
the latest observation of cosmic microwave background radiation
(CMBR). Although the single field inflation model has obtained
fruitful achievements, we still need to explore more on this theory.

A significant lesson is to investigate its higher order
perturbations. There has been a number of literature studying the
behavior of higher order perturbations in inflation models
(see~\cite{NG} for an excellent pioneer work, and see~\cite{bartolo}
for a good review on this issue). From the viewpoint of statistic
dynamics, these higher order perturbations are usually related to
$n$ ($n>2$) point correlators. So if these correlators indeed exist,
there must be non-Gaussianity in the early universe. The
non-Gaussianity is a very important issue worth studing. Since the
non-Gaussianity has many features which can be observed by
experiments, such as its magnitude, shape, running and so on, it
encodes plentiful information about the early universe. We are able
to learn what has happened since that time if we detect it. For
example, we have already known that the primordial bispectrum of
single scalar field inflation model is too small to be
observed~\cite{NG}. However, there are implications of
non-Gaussianity which value may be large from astronomical
data~\cite{wmap,pfnl} recently. If this is confirmed, the usual
inflation models, especially chaotic inflation, will suffer a great
challenge from the experiments. Another example is, that different
models usually have different predictions about
non-Gaussianity~\cite{bartolo,arkani,DBI,K,lyth-rodriguez,lyth-rodriguez-a,kejie,chen,thermal,yi},
and thus the non-Gaussianity may help us to discriminate these
models.

One may be interested in how to produce large non-Gaussianity in
inflation. Firstly, let us think about why the non-Gaussianity of
single scalar field inflation model is so small. Ordinarily, the
non-Gaussian effect comes from the interactions of the perturbation
variables in the canonical case that we choose the Bunch-Davis
vacuum. In the single scalar field inflation model, the interaction
terms of the perturbation variables are strongly suppressed by slow
roll parameters. However, if we modify the lagrangian of the
inflaton to be non-canonical, such as DBI inflation~\cite{DBI,chen}
and K-inflation~\cite{K}, it is possible to obtain a large value of
non-Gaussianity. Another way to obtain large bispectrum is to change
the initial condition of Gaussian statistic, e.g. a thermal initial
condition~\cite{thermal}.

The above arguments are valid when we only consider the adiabatic
perturbations. It is feasible since the perturbations generated in
single scalar field inflation model are always highly adiabatic and
the curvature perturbation $\zeta$ is conserved on large scale.
However, this picture is changed when the inflation is driven by
multiple fields~\cite{enqvist-jokinen-b,enqvist-vaihkonen}. When we
introduce multiple fields, they will generate a large amount of
entropy fluctuations which are converted into the curvature
perturbations at later time. This scenario can result in large
non-Gaussianity of local form. For example, the curvaton
mechanism~\cite{lyth-wands,curvaton} and the in-homogenous reheating
scenario~\cite{Dvali:2003em,Dvali:2003ar,Zaldarriaga:2003my} are
able to produce large non-Gaussianity of local form. Therefore, it
is meaningful to study the bispectrum of multiple-field inflation
with both the magnitude and the shape in detail.

Interestingly, the method of calculating the primordial
non-Gaussianity is not unique, and a number of methods have been
proposed in literature. These methods possess different advantages
in different occasions. The most direct formalism was developed
by~\cite{NG2} in which the authors calculated the second order
perturbations from the Einstein equations; another useful formalism
was called Lagrangian formalism~\cite{NG} which derived the
interaction terms of curvature perturbations in the Lagrangian. Both
the two approaches are able to calculate the magnitude and the shape
of the non-Gaussianity of which the local one is the most interested
in observations. Moreover, a so-called $\delta N$
formalism~\cite{deltaN} has been proposed to calculate the local
non-Gaussianity specifically. This method greatly simplified the
calculation of non-Gaussianity.
Some pioneer works on the non-Gaussianity of multiple-field
inflation have been done by using different methods. For example,
the local form of non-Gaussianity in two-field inflation is shown
by~\cite{two ng} based on $\delta N$ formalism; the shape of
non-Gaussianity in multiple-field inflation is given by the
Lagrangian formalism in~\cite{multi ng}. Recently, a remarkable work
has been done by~\cite{eom} in which the authors have used
second-order Klein-Gordon equation~\cite{kg,kg2} to calculate the
non-Gaussianity, which is consistent with the Lagrangian formalism.
The method of field equation can directly derive the non-Gaussianity
from equation of motion, without assuming an effective action
principle. We in this paper extend the field equation formalism to
the multiple-field inflation and calculate the non-Gaussianity. In
the derivation, we assume that there are $\mathcal N$ scalar fields
$\phi^{I},\ \phi^{J}, \ \cdots$ in the period of inflation, and the
potential $V$ of the scalar fields depends on them. We take the
natural unit $M_P\equiv(8\pi G)^{-1/2}=1$ in this paper.



Our paper is organized as follows. In Section \S\ref{background}, we
review the background evolution of the multiple-field inflation, and
define the slow-roll parameters. In section \S\ref{first order}, the
quantum theory of the first order perturbations is discussed. By
means of the canonical method, we quantize the perturbations of
scalar fields, and present the Green's functions. In Section
\S\ref{KG}, we derive the second order Klein-Gordon equation
directly from the action, and so the second order perturbations of
scalar fields are obtained by the Green's function. Section
\S\ref{correlator} presents the main result of our paper which shows
that there are different source terms contributing to the
three-point correlator of scalar fluctuations. In Section
\S\ref{ng}, we review the $\delta N$ formalism, and calculate the
nonlinear parameter $f_{\rm NL}$. Conclusions and discussions are
summarized in the last section.

\section{The background in multiple-field inflation}
\label{background}

In this section, we show the field equations in the background, and
define some slow roll parameters in multiple-field inflation. The
background is assumed to be the Friedmann-Robertson-Walker (FRW)
spacetime, and the action takes the form \be \d s^2 \, = \, -\d t^2
+a(t)^2 \delta_{ij} \, \d x^i \, \d x^j ~ , \ee where $a(t)$ is the
scale factor. In some cases, it is convenient to use conformal time
$\eta$, which is defined as $\eta \equiv \int^{\infty}_t \d
t^{\prime}/ a(t^{\prime})$. And to the leading order of slow roll
approximation, $\eta \sim -\frac{1}{a H}$ in the period of
inflation.

The equation of scalar field takes the form \be \phi_0^{I \, \prime
\prime}
 + 2 \H \phi_0^{I \, \prime}+V_{,\,I}=0\ , \ee
where $I$ denotes different scalar fields, prime denotes
$\frac{\d}{\d \, \eta}$, $V_{,\,I}$ is the shorthand for $\frac{\d
V}{\d \phi^I}$, $\H \equiv a^{\prime}/a$ is the conformal Hubble
scale, and the metric of the field space is assumed to be
$\delta_{IJ}$.

The 0-0 component of Einstein equations gives the so-called
Friedmann equation, \be 3 \H^2= \frac{1}{2} \delta_{IJ} \phi_0^{I \,
\prime} \phi_0^{J \, \prime}+ a^{2} V(\phi_0)\ . \ee And from the
i-j component of Einstein equations, we have \be
\H^2+2\H^{\prime}=-\frac{1}{2} \delta_{IJ} \phi_0^{I \, \prime}
\phi_0^{J \, \prime}+a^{2} V(\phi_0)\ , \ee where the repeated up
index and down index represent summation.

As in single field inflation, the potential should satisfy the slow
roll condition due to the constraint from the observation. It
requires that the velocity and acceleration of inflaton rolling down
the potential are very small. In the multiple-field inflation, we
use the slow roll matrix \be \epsilon^{IJ}=\frac{\dot \phi_0^{I}
\dot \phi_0^{J}}{2 H^2}=\frac{\phi_0^{I \, \prime} \phi_0^{J \,
\prime}}{2 \H^2}= \epsilon^{I} \epsilon^{J}, \ee where \be
\label{epsilon} \epsilon^{I}=\frac{\dot \phi_0^{I}}{\sqrt{2} H} \
,\ee and the trace of the slow roll matrix $\mathrm{tr}
\,\epsilon^{IJ}$ is the standard slow roll parameter $\epsilon =
-\dot H/H^2$. In general situation, the order of these slow roll
parameters are estimated as \be \epsilon^{IJ} \sim \mathcal{O}
(\frac{\epsilon}{\mathcal{N}}), \ \epsilon^{I} \sim \mathcal{O}
(\sqrt{\frac{\epsilon}{\mathcal{N}}})\ . \ee To generalize the
single field inflation, we introduce the second slow roll matrix,
\be \eta^{IJ} = \frac{\ddot{\phi}^I \dot{\phi}^J + \dot{\phi}^J
\ddot{\phi}^J} {4 H \dot{H}} \ . \ee The diagonal element of this
matrix is the slow parameter in the single field inflation
$\eta^{\phi\phi}=-\frac{\ddot \phi}{H \dot \phi}=\eta$.

\section{The First-order perturbation in the uniform curvature gauge}
\label{first order}

When we compute the perturbation of inflation, the quantity is
usually changed with the coordinate transformation. In order to
discuss the real physical freedoms in the inflationary perturbation
theory, we should select a gauge~\cite{bran}. Fixing a gauge means
choosing a coordinate system. Different gauges are equivalent in
physics. In this section we select the uniform curvature gauge and
discuss the first order perturbation of real physical freedoms. It
is convenient to study in ADM formalism and the metric can be
expressed as \be \d \, s^2 = - \mathrm N^2 \d t^2 + h_{ij} (\d x^i +
\mathrm N^i dt)(\d x^{j} +\mathrm N^j \d t), \ee so the action is
\bqa \label{action} S &=& - \frac{1}{2} \int \mathrm N \sqrt{h} \,
\left( \delta_{IJ} h^{ij}
\partial_i \phi^I
  \partial_j \phi^J - 2 V(\phi) \right)\nonumber\\
   &&+
  \frac{1}{2} \int \mathrm N^{-1} \sqrt{h} \, \left( E_{ij} E^{ij} - E^2 +
  \delta_{IJ} (\dot{\phi}^I - \mathrm N^j \partial_j \phi^I) (\dot{\phi}^J
   - \mathrm N^j \partial_j \phi^J) \right) \ ,
\eqa where $\mathrm N^{-1}E_{ij}$ is the extrinsic curvature,
$E=E^i_i$. We select the uniform curvature gauge, in which the Ricci
curvature is zero at the same coordinate $t$ and
$h_{ij}=a^2(t)\delta_{ij}$. The two scalar perturbations from the
metric perturbation can be expressed by the lapse $\mathrm N$, and
shift $\mathrm N^{i}$. The lapse $\mathrm N$, and shift $\mathrm
N^{i}$ are Lagrangian multipliers. Thus the physical freedoms can be
expressed by the $\mathcal N$ scalar perturbations $\delta \phi^{I}$
in the uniform curvature gauge.

As in the single field, the scalar perturbation can be expanded in
powers of the gaussian perturbation $\delta \phi^{I}_1$, \be
\delta\phi^{I} = \delta\phi^I_1 + \frac{1}{2} \delta\phi^I_2 +
\cdots + \frac{1}{n!} \delta\phi^I_n + \cdots . \ee The closer the
primordial scalar perturbation is to gaussian statistics, the better
the expansion is.

Since $\delta \phi^{I}_1$ obeys the gaussian statistics, the
equation of motion of $\delta \phi^{I}_1$ is linear. After some
simplification of (\ref{action}), the second order action takes the
form \be \label{S2} S_2 = \frac{1}{2} \int \d \eta \, \d^ {3}{x} \;
a^2 \left( \delta_{IJ} \delta\phi^{I\, \prime}_1 \delta\phi^{J\,
\prime}_1 - \delta_{IJ} \partial \delta\phi^I_1 \partial
\delta\phi^J_1 \right)\  , \ee where $\partial \delta\phi^I_1
\partial \delta\phi^J_1$ is the shorthand for the scalar
product $\delta^{ij} \partial_i \delta\phi^I_1 \partial_j
\delta\phi^J_1$. Then the field equation of scalar field $\delta
\phi^{I}$ for the Fourier mode is \be \delta \phi^{I\,
\prime\prime}_1 + 2 \H \delta \phi^{I\, \prime}_1 + k^2 \delta
\phi^{I}_1 = 0 ,\ee The classical field is quantized by the
canonical method, \be
 \label{aa}
        \delta\hat{\phi}^{I}_1({\bf{x}},\eta) = \int \frac{\d^3 k}{(2\pi)^3}
       e^
        { i \bf{k}\cdot\bf{x}}
        \{
            a^{I \dag}_{\bf{k}} \theta^{I}_k(\eta) +
            a^I_{-\bf{k}} \bar{\theta}^{I}_k(\eta)
       \} \ ,
 \ee
where $\theta^{I}_k$, $\bar{\theta}^I_{k}$ are massless scalar
fields in momentum space. The normalization of the terms is
determined by the commutative relation between scalar field and its
canonical momentum, and the commutative relation between the
creation and annihilation operator \be \label{a} [a^I_{\bf{k}}, a^{J
\dag}_{\bf{k}^\prime}]=(2 \pi)^3 \delta^{IJ}
\delta(\bf{k}-\bf{k}^\prime) \ .\ee

In the Bunch-Davies vacuum, the normalized scalar field
is~\cite{qft},
 \be
\label{theta}
  \theta^I_k =\frac{H}{\sqrt{2 k^3}}(1- i k \eta)
e^{ik \eta} \ .\ee Since the value of $\theta^I_k$ is independent of
$I$, we omit the index $I$ in $\theta^I_k$ afterwards. The two-point
correlator of scalar fields is \bqa
        \langle \delta\phi^I_1({\bf{k}},\eta) \delta\phi^J_1({\bf{k}'},\eta')
        \rangle &=& (2 \pi)^3 \delta^{IJ} \delta({\bf{k}} + {\bf{k}'}) \bar{\theta}_{k}(\eta)
        \theta_{k}(\eta')\nonumber\\
         &\sim & (2 \pi)^3 \delta^{IJ} \delta({\bf{k}} + {\bf{k}'})\frac{H^2}{2 k^3} \ \ \ \ \ \ \ \ \  for
         \ \  \ k\eta \ll 1  \nonumber\\
         &=&(2 \pi)^3 \delta^{IJ} \delta({\bf{k}} +
         {\bf{k}'})\frac{2\pi^2}{k^3} \mathrm P(k)\ ,
         \label{P}
    \eqa
where $\mathrm P(k)=\frac{H^2}{4 \pi^2}$ is the so-called power
spectrum of scalar field. And the retarded Green's function in
momentum space takes the form, \be
        Gr_k(\eta,\tau) = i a(\tau)^2 \times \left\{
            \begin{array}{l@{\hspace{5mm}}l}
                0   & \eta < \tau \\
                \theta_k(\tau) \bar{\theta}_k(\eta) -
                \bar{\theta}_k(\tau) \theta_k(\eta) & \eta > \tau
            \end{array} \right. .
    \ee
Using the Green's function, the second order field equation can be
solved as a linear function with the source term.

\section{The second-order Klein-Gordon equation}
\label{KG}

In this section, we derive the second-order Klein-Gordon equation
from the multiple-field action (\ref{action}). The situation of
single field is given by~\cite{kg2}. Expanding the action
(\ref{action}), it includes the terms of all the scalar fields and
scalar perturbations from the metric. The lapse and the shift in the
action are determined since they are Lagrangian multipliers without
dynamics effect. Finally they are eliminated from the action which
only contains the second-order perturbation of scalar fields $\delta
\phi^I_2$. The part of the action quadratic in $\delta \phi^I_2$ can
be expressed in conformal time, \be S_2 = \frac{1}{8} \int \d \eta
\, \d^ {3}{x} \; a^2 \left( \delta_{IJ} \delta\phi^{I\, \prime}_2
\delta\phi^{J\, \prime}_2 - \delta_{IJ} \partial \delta\phi^I_2
\partial \delta\phi^J_2 \right)\ . \ee and the cubic term in the
slow roll approximation~\cite{multi ng} is \bqa S_3 &=& \int \d \eta
\d ^3 x a^2 [\frac{1}{3!} V_{,\,IJK}\delta \phi^{I}_2 \delta
\phi^{J}_1 \delta \phi^{K}_1+\delta_{IJ}\delta_{MN} \frac{\delta
\phi^{M \, \prime}_0}{4\H} \delta \phi^{I\, \prime}_2
\partial \nabla^{-2}(\delta \phi^{N\, \prime}_1) \partial\delta \phi^{J}_1 -
\nonumber\\
&& \delta_{IJ}\delta_{MN} \frac{\delta \phi^{M \, \prime}_0}{8\H}
\delta \phi^{N}_2 \delta \phi^{I \, \prime}_1 \phi^{J \, \prime}_1 -
\delta_{IJ}\delta_{MN} \frac{\delta \phi^{M \, \prime}_0}{8\H}
\delta \phi^{N}_2 \partial \delta \phi^{I}_1
\partial \delta \phi^{J}_1] + perms \ , \eqa
where the permutations represent swapping the $\delta \phi_2$ in
other possible positions. The term containing $V_{,\,IJK}$ in the
action is not neglected by the slow-roll approximation, because it
may contribute large effect in
non-Gaussianity~\cite{Zaldarriaga:2003my}.

Variation $\delta S / \delta (\delta \phi^{I}_2)=0$ gives the field
equation. All the surface terms are neglected, which requires that
$\delta(\delta \phi^{I}_1)$ vanishes in the boundary, and the
equation of motion for $\delta \phi^{I}_1$ simplifies the result
further. The final result is \bqa &&\delta \phi^{I\, \prime\prime}_2
+ 2 \H
\delta \phi^{I\, \prime}_2  + k^2 \delta \phi^{I}_2\nonumber\\
 &=&
(-a^2 V_{,\,IJK} \delta \phi^{J}_1 \delta \phi^{K}_1 ) +\frac{\delta
\phi^{M \, \prime}_0}{\H} (-2\delta_{MN}\partial \nabla^{-2} \delta
\phi^{N\, \prime}_1
\partial \delta \phi^{I \, \prime}_1+ 2\delta_{MN} \delta \phi^{N}_1
\nabla^{2} \delta
\phi^{I}_1 ) \nonumber\\
&& +\frac{\delta \phi^{I \, \prime}_0}{\H} [-\frac{1}{2} \delta_{MN}
\delta \phi^{M\, \prime}_1  \delta \phi^{N \, \prime}_1 -\frac{1}{2}
\delta_{MN} \partial \delta \phi^{M}_1 \partial \delta \phi^{N}_1
+\delta_{MN}\nabla^{-2}(\partial \nabla^{2}\delta \phi^{M}_1
\partial \delta \phi^{N}_1\nonumber\\
&& +\nabla^{2}\delta \phi^{M}_1 \nabla^{2} \delta \phi^{N}_1 +\delta
\phi^{M\, \prime}_1 \nabla^{2} \delta \phi^{N \,
\prime}_1+\partial\delta \phi^{M\, \prime}_1 \partial \delta \phi^{N
\, \prime}_1)] \ . \eqa
On the right hand side of the equation it is the source term. Using
the Green's function, $\delta \phi^{I}_2$ takes the form \be
\label{phi2} \delta\phi^{I}_2(\eta,{\bf{x}}) =
            \int \frac{\d^3 q}{(2\pi)^3} e^{i {\bf{q}}\cdot{\bf{x}}}
            \left\{
                \int_{-\infty}^{\eta} \d \tau
                \int \frac{\d ^3 k_1 \, \d ^3 k_2}{(2\pi)^6}
                Gr_q(\eta,\tau)
                \delta({\bf{q}} - {\bf{k}_1} - {\bf{k}_2}) \mathcal S
            \right\} ,
 \ee
where \bqa \mathcal S &\equiv& -a^2 V_{,\,IJK} \delta \phi^{J}_1
\delta \phi^{K}_1+\delta_{MN} \mathcal F_1 \frac{\delta \phi^{M \,
\prime}_0}{\H}\delta \phi^{N}_1 \delta \phi^{I}_1+ \delta_{MN}
\mathcal F_2 \frac{\delta \phi^{I \, \prime}_0}{\H}\delta \phi^{M}_1
\delta \phi^{N}_1 \nonumber\\
&&+\delta_{MN} \mathcal G_1 \frac{\delta \phi^{M \,
\prime}_0}{\H}\delta \phi^{N\, \prime}_1 \delta \phi^{I \,
\prime}_1+\delta_{MN} \mathcal G_2 \frac{\delta \phi^{I \,
\prime}_0}{\H}\delta \phi^{M \, \prime}_1 \delta \phi^{N \,
\prime}_1, \eqa and $\{\mathcal F_1, \mathcal F_2, \mathcal G_1,
\mathcal G_2\}$ are some factors in the momentum space. \bqa
&&\mathcal F_1=-2 k_2^2\ ,\ \  \mathcal F_2=
             \frac{1}{2} {\bf{k}_1} \cdot {\bf{k}_2} -
            \frac{1}{({\bf{k}_1} + {\bf{k}_2})^2} \left(
                k_1^2 k_2^2 + k_1^2 {\bf{k}_1} \cdot {\bf{k}_2}
                \right)
            \\
&&\mathcal G_1=- \frac{2}{k_1^2} {\bf{k}_1} \cdot {\bf{k}_2} \ , \ \
     \mathcal G_2=-       \frac{1}{2} +
            \frac{1}{(\bf{k}_1 + \bf{k}_2)^2} \left(
                k_2^2 + {\textbf{k}_1} \cdot {\bf{k}_2} \right) \eqa
Notice that the terms with $\mathcal F_2$ and $\mathcal G_2$ are
symmetric with $M,N$, so when we calculate the three-point function,
$\mathcal F_2$ and $\mathcal G_2$ must be symmetrized over
permutations of $\{k_1,k_2\}$ as in~\cite{eom}. On the other hand,
$\mathcal F_1$ and $\mathcal G_1$ cannot be symmetrized.

\section{Three-point correlator}
\label{correlator}

The three-point correlator of a free scalar field vanishes, $\langle
\delta\phi_1 \delta\phi_1 \delta\phi_1
        \rangle=0$. The leading order of three-point correlator $\langle
\delta\phi \delta\phi \delta\phi
        \rangle$ is $\langle
\delta\phi_1 \delta\phi_1 \delta\phi_2
        \rangle \sim \frac{1}{2}\langle
\delta\phi_1 \delta\phi_1 \delta\phi_1 \ast \delta\phi_1
        \rangle $, where $\ast$ denotes a convolution. Thus with the
value of $\delta \phi^{I}_1$ (\ref{aa}) and $\delta \phi^{I}_2$
(\ref{phi2}), the three-point correlator of multiple-field can be
calculated. As the argument given in~\cite{eom}, the field equation
of multiple-field is in the approximation of slow roll limit, and
the expansion in powers of slow-roll parameter is not applicable at
the end of inflation. The reason is that the the subleading term has
logarithmic divergences $\mathrm{ln}|k\eta|=N$, and the growth of
the e-folding number makes the subleading terms not negligible. Here
we just calculate the three-point correlator when the modes cross
the horizon.

According to the source term of the field equation, we will show the
results of three-point correlator from the three parts below.

\subsection{$V_{,\,IJK}$ terms}

In slow roll approximation, we neglect the $V_{,\,I}$ and
$V_{,\,IJ}$ terms, but the $V_{,\,IJK}$ terms could have
non-neglectable effect in some situation, and also lead to the
logarithmic divergence.

The three-point correlator from $V_{,\,IJK}$ terms take the form
 \bqa
        \langle \delta\phi^{I}(\textbf{k}_1) \delta\phi^J(\textbf{k}_2)
        \delta\phi^K(\textbf{k}_3) \rangle & \supseteq &
        - i (2\pi)^3 \delta(\textbf{k}_1 + \textbf{k}_2 + \textbf{k}_3)
        \int_{-\infty}^{\eta} \d\tau \; a(\tau)^4 V_{,\,IJK} \times  \nonumber\\
        & & \hspace{-2.1cm}
        \Bigg\{ \left[
            \theta_{k_3}(\tau)\bar{\theta}_{k_3}(\eta) -
            \bar{\theta}_{k_3}(\tau)\theta_{k_3}(\eta) \right]
        \bar{\theta}_{k_1}(\eta) \bar{\theta}_{k_2}(\eta)
        \theta_{k_1}(\tau) \theta_{k_2}(\tau) + \mbox{}  \nonumber\\
        & & \hspace{-1.7cm}
        \left[
            \theta_{k_2}(\tau)\bar{\theta}_{k_2}(\eta) -
            \bar{\theta}_{k_2}(\tau)\theta_{k_2}(\eta) \right]
        \bar{\theta}_{k_1}(\eta) \theta_{k_3}(\eta)
        \theta_{k_1}(\tau) \bar{\theta}_{k_3}(\tau) + \mbox{}  \nonumber\\
        & & \hspace{-1.7cm}
        \left[
            \theta_{k_1}(\tau)\bar{\theta}_{k_1}(\eta) -
            \bar{\theta}_{k_1}(\tau)\theta_{k_1}(\eta) \right]
        \theta_{k_2}(\eta) \theta_{k_3}(\eta)
        \bar{\theta}_{k_2}(\tau) \bar{\theta}_{k_3}(\tau) \Bigg\} .
        \label{Vterm}
    \eqa
The $\theta$ terms in the brackets come from the Green's function.
Outside the brackets, the $\theta$ terms which depend on time
parameter $\eta$ are derived from the free scalar field, and the
$\theta$ terms depending on $\tau$ is derived from the source of the
field equation. We have stated that $\theta$ is independent of the
index in multiple-field, so the final result is similar to single
field for $V_{,\,IJK}$ term. With the value of $\theta$
(\ref{theta}), this part of three-point correlator can be written to
the leading order of slow roll approximation, \bqa \langle
\delta\phi^{I}(\textbf{k}_1) \delta\phi^J(\textbf{k}_2)
        \delta\phi^K(\textbf{k}_3) \rangle  &\supseteq&  (2\pi)^3
\delta(\textbf{k}_1 + \textbf{k}_2 + \textbf{k}_3)
        \frac{H_\ast^2 V_{\ast,\,IJK}}{4 \prod_i k_i^3} \times
        \nonumber\\
&&\hspace{-2.3cm}\int_{-\infty}^{\eta} \frac{\d\tau}{\tau^4} \;
\mathrm{Re}[-i(1-ik_1\tau)(1-ik_2\tau)(1-ik_3\tau)e^{ik_t \tau}]\ ,
        \eqa
where $k_t=k_1+k_2+k_3$, $\ast$ denotes the value at the time
$\eta$. Here we take $\eta$ to the value that the modes cross the
horizon. Following the paper~\cite{NG}, we can deform the
integration variable $\tau$ to Euclidean time and deal with the
divergence of the integral properly. Finally, we obtain \bqa \langle
\delta\phi^{I}(\textbf{k}_1) \delta\phi^J(\textbf{k}_2)
        \delta\phi^K(\textbf{k}_3) \rangle  &\supseteq&  (2\pi)^3
\delta(\textbf{k}_1 + \textbf{k}_2 + \textbf{k}_3)
        \frac{H_\ast^2 V_{\ast,IJK}}{4 \prod_i k_i^3} \times
        \nonumber\\
        &&\hspace{-2.3cm} \left(
            - \frac{4}{9} k_t^3 + k_t \prod_{i < j} k_i k_j +
            \frac{1}{3} \Big\{ \frac{1}{3} + \gamma +
                \ln | k_t \eta | \Big\} \sum_i k_i^3
        \right) \ ,
        \eqa
where $i\in \{ 1, 2, 3 \}$, and $\gamma \approx 0.577$ is the
Euler's constant. In multiple-field, there also exists the infra-red
divergence term form the $V_{,\,IJK}$ terms. When we take the time
crossing the horizon, the divergence term is negligible. The
classical evolution of perturbation afterwards will make the term
large, but we can use other formalism to deal with the problem, such
as $\delta N$ formalism, or the separate universe
approach~\cite{star,conservation of eta}.

\subsection{$\mathcal F$ terms}
In this section, we discuss the zero-derivative terms in the source,
which contain the contributions of the $\mathcal F_1$ and $\mathcal
F_2$ terms. Since there is a delta function in the three-point
correlator, the sum of momentum $\textbf{k}_1 + \textbf{k}_2 +
\textbf{k}_3$ is zero, we could write $\mathcal F_1$ and $\mathcal
F_2$ in other equivalent form, \be \mathcal
        F_1(k_1,k_2;k_3)=-2k_2^2\ , \ee
and $\mathcal F_2$ could be symmetrized as \be \mathcal
F_2(k_1,k_2;k_3)= -\frac{1}{2} (k_1^2 + k_2^2) +\frac{(k_1^2 -
k_2^2)^2}{4 k_3^2} + \frac{k_3^2}{4} \ .\ee The part of three-point
correlator come from the $\mathcal F_1$ and $\mathcal F_2$ terms is
expressed as \bqa \langle \delta\phi^{I}(\textbf{k}_1)
\delta\phi^J(\textbf{k}_2)
        \delta\phi^K(\textbf{k}_3) \rangle & \supseteq &
         i (2\pi)^3 \delta(\textbf{k}_1 + \textbf{k}_2 + \textbf{k}_3)
        \int_{-\infty}^{\eta} \d\tau \; a(\tau)^2 \times \frac{1}{2} \times\nonumber\\
        \nonumber & & \hspace{-4.3cm}
        \Bigg\{ [\sqrt{2}\epsilon^{I}\delta^{JK} \mathcal F_1(k_1,k_2;k_3)+
        \sqrt{2}\epsilon^{J}\delta^{IK} \mathcal F_1(k_2,k_1;k_3)+
        2\sqrt{2}\epsilon^{K}\delta^{IJ} \mathcal F_2(k_1,k_2;k_3)]
        \\
        \nonumber & & \hspace{-4cm}
        \left[
            \theta_{k_3}(\tau)\bar{\theta}_{k_3}(\eta) -
            \bar{\theta}_{k_3}(\tau)\theta_{k_3}(\eta) \right]
        \bar{\theta}_{k_1}(\eta) \bar{\theta}_{k_2}(\eta)
        \theta_{k_1}(\tau) \theta_{k_2}(\tau) + \mbox{} \\
        \nonumber & & \hspace{-4cm}
        \mathcal [\sqrt{2}\epsilon^{I}\delta^{JK} \mathcal F_1(k_1,k_3;k_2)+
        \sqrt{2}\epsilon^{K}\delta^{IJ} \mathcal F_1(k_3,k_1;k_2)+
        2\sqrt{2}\epsilon^{J}\delta^{IK} \mathcal F_2(k_3,k_1;k_2)]
        \\
        \nonumber & & \hspace{-4cm} \left[
            \theta_{k_2}(\tau)\bar{\theta}_{k_2}(\eta) -
            \bar{\theta}_{k_2}(\tau)\theta_{k_2}(\eta) \right]
        \bar{\theta}_{k_1}(\eta) \theta_{k_3}(\eta)
        \theta_{k_1}(\tau) \bar{\theta}_{k_3}(\tau) + \mbox{} \\
        & & \hspace{-4cm}
        [\sqrt{2}\epsilon^{J}\delta^{IK} \mathcal F_1(k_2,k_3;k_1)+
        \sqrt{2}\epsilon^{K}\delta^{IJ} \mathcal F_1(k_3,k_2;k_1)+
        2\sqrt{2}\epsilon^{I}\delta^{JK} \mathcal F_2(k_2,k_3;k_1)]
        \nonumber\\
         & & \hspace{-4cm}
        \left[
            \theta_{k_1}(\tau)\bar{\theta}_{k_1}(\eta) -
            \bar{\theta}_{k_1}(\tau)\theta_{k_1}(\eta) \right]
        \theta_{k_2}(\eta) \theta_{k_3}(\eta)
        \bar{\theta}_{k_2}(\tau) \bar{\theta}_{k_3}(\tau) \Bigg\} ,
\eqa where $\delta^{IJ}$ origins from the commutation relation of
creation and annihilation operators (\ref{a}), and the factor $1/2$
ahead of the open brace comes from the definition of
$\delta\phi^{I}_2$. Different modes of contracting a free scalar
field $\delta\phi_1$ and $\delta\phi_1$ in the source term make
$\mathcal F_2$ symmetric, and cause the factor $2$ in front of
$\mathcal F_2$. Here the slow-roll parameter $\epsilon^{I}$ is
defined in (\ref{epsilon}).

In the three-point correlator, $\mathcal F$ terms are independent of
integration variable, so the factors of final results $f_1$, $f_2$,
and $f_3$ are similar to the case of single field, \bqa \langle
\delta\phi^{I}(\textbf{k}_1) \delta\phi^J(\textbf{k}_2)
        \delta\phi^K(\textbf{k}_3) \rangle & \supseteq &
          (2\pi)^3 \delta(\textbf{k}_1 + \textbf{k}_2 + \textbf{k}_3)\frac{H_\ast^4}{8 \prod_i
          k_i^3} \times \frac{1}{2} \nonumber\\
          &&\hspace{-5.6cm}\Big\{ f_1  [\sqrt{2}\epsilon^{I}\delta^{JK} \mathcal F_1(k_1,k_2;k_3)+
        \sqrt{2}\epsilon^{J}\delta^{IK} \mathcal F_1(k_2,k_1;k_3)+
        2\sqrt{2}\epsilon^{K}\delta^{IJ} \mathcal F_2(k_1,k_2;k_3)]
        + \nonumber\\
&&\hspace{-5.3cm}
            f_2 [\sqrt{2}\epsilon^{I}\delta^{JK} \mathcal F_1(k_1,k_3;k_2)+
        \sqrt{2}\epsilon^{K}\delta^{IJ} \mathcal F_1(k_3,k_1;k_2)+
        2\sqrt{2}\epsilon^{J}\delta^{IK} \mathcal F_2(k_3,k_1;k_2)]+
        \nonumber\\
  &&\hspace{-5.3cm}
            f_3 [\sqrt{2}\epsilon^{J}\delta^{IK} \mathcal F_1(k_2,k_3;k_1)+
        \sqrt{2}\epsilon^{K}\delta^{IJ} \mathcal F_1(k_3,k_2;k_1)+
        2\sqrt{2}\epsilon^{I}\delta^{JK} \mathcal F_2(k_2,k_3;k_1)]
        \Big\},
\eqa where \bqa
        f_1 & \equiv & - \frac{2 k_3^3(k_1^2 + 4 k_1 k_2 + k_2^2 - k_3^2)}
                                {(k_1 + k_2 - k_3)^2 k_t^2}\,, \label{eq:fa}\nonumber \\
        f_2 & \equiv & - \frac{2 k_2^3(k_1^2 - 4 k_1 k_3 + k_3^2 - k_2^2)}
                                {(k_2^2 - (k_1 - k_3)^2)^2}\,, \label{eq:fb}\nonumber \\
        f_3 & \equiv & - \frac{2 k_1^3(k_2^2 + 4 k_2 k_3 + k_3^2 - k_1^2)}
                                {(k_1 - k_2 - k_3)^2 k_t^2}\,.\label{eq:fc}
    \eqa
The factor $H^{4}_\ast$ origins from $\theta$ and the scale factor
$a$.

\subsection{$\mathcal G$ terms}

Similar to the derivation of $\mathcal F$ terms, $\mathcal G$ terms
lead to another part of the three-point correlator. We have \be
\mathcal G_1(k_1,k_2,k_3)=\frac{k_1^2 + k_2^2 - k_3^2}{k_1^2} \ ,\ee
and $\mathcal G_2=0$ when symmetrized. The expectation value of
$\mathcal G$ terms is \bqa \langle \delta\phi^{I}(\textbf{k}_1)
\delta\phi^J(\textbf{k}_2)
        \delta\phi^K(\textbf{k}_3) \rangle & \supseteq &
         i (2\pi)^3 \delta(\textbf{k}_1 + \textbf{k}_2 + \textbf{k}_3)
        \int_{-\infty}^{\eta} \d\tau \; a(\tau)^2 \times \frac{1}{2}\times \nonumber\\
        \nonumber & & \hspace{-4.3cm}
        \{ [\sqrt{2}\epsilon^{I}\delta^{JK} \mathcal G_1(k_1,k_2;k_3)+
        \sqrt{2}\epsilon^{J}\delta^{IK} \mathcal G_1(k_2,k_1;k_3)
        ]
        \\
        \nonumber & & \hspace{-4cm}
        \left[
            \theta_{k_3}(\tau)\bar{\theta}_{k_3}(\eta) -
            \bar{\theta}_{k_3}(\tau)\theta_{k_3}(\eta) \right]
        \bar{\theta}_{k_1}(\eta) \bar{\theta}_{k_2}(\eta)
         \frac{\d}{\d\tau}\theta_{k_1}(\tau) \frac{\d}{\d\tau}\theta_{k_2}(\tau) + \mbox{} \\
        \nonumber & & \hspace{-4cm}
        \mathcal [\sqrt{2}\epsilon^{I}\delta^{JK} \mathcal G_1(k_1,k_3;k_2)+
        \sqrt{2}\epsilon^{K}\delta^{IJ} \mathcal
        G_1(k_3,k_1;k_2)]
        \\
        \nonumber & & \hspace{-4cm} \left[
            \theta_{k_2}(\tau)\bar{\theta}_{k_2}(\eta) -
            \bar{\theta}_{k_2}(\tau)\theta_{k_2}(\eta) \right]
        \bar{\theta}_{k_1}(\eta) \theta_{k_3}(\eta)
       \frac{\d}{\d\tau} \theta_{k_1}(\tau) \frac{\d}{\d\tau}\bar{\theta}_{k_3}(\tau) + \mbox{} \\
        & & \hspace{-4cm}
        [\sqrt{2}\epsilon^{J}\delta^{IK} \mathcal G_1(k_2,k_3;k_1)+
        \sqrt{2}\epsilon^{K}\delta^{IJ} \mathcal G_1(k_3,k_2;k_1)]
        \nonumber\\
         & & \hspace{-4cm}
        \left[
            \theta_{k_1}(\tau)\bar{\theta}_{k_1}(\eta) -
            \bar{\theta}_{k_1}(\tau)\theta_{k_1}(\eta) \right]
        \theta_{k_2}(\eta) \theta_{k_3}(\eta)
        \frac{\d}{\d\tau}\bar{\theta}_{k_2}(\tau) \frac{\d}{\d\tau}\bar{\theta}_{k_3}(\tau) \}
        .
\eqa Repeating the progress in the case of $\mathcal F$ terms, we
obtain \bqa \langle \delta\phi^{I}(\textbf{k}_1)
\delta\phi^J(\textbf{k}_2)
        \delta\phi^K(\textbf{k}_3) \rangle & \supseteq &
          (2\pi)^3 \delta(\textbf{k}_1 + \textbf{k}_2 + \textbf{k}_3)\frac{H_\ast^4}{8 \prod_i
          k_i^3} \times \frac{1}{2}\times
          \nonumber\\
          &&\hspace{-5.3cm}\Big\{ g_1  [\sqrt{2}\epsilon^{I}\delta^{JK} \mathcal G_1(k_1,k_2;k_3)+
        \sqrt{2}\epsilon^{J}\delta^{IK} \mathcal G_1(k_2,k_1;k_3)
        ]
        + \nonumber\\
&&\hspace{-5cm}
            g_2 [\sqrt{2}\epsilon^{I}\delta^{JK} \mathcal G_1(k_1,k_3;k_2)+
        \sqrt{2}\epsilon^{K}\delta^{IJ} \mathcal
        G_1(k_3,k_1;k_2)]+
        \nonumber\\
  &&\hspace{-5cm}
            g_3 [\sqrt{2}\epsilon^{J}\delta^{IK} \mathcal G_1(k_2,k_3;k_1)+
        \sqrt{2}\epsilon^{K}\delta^{IJ} \mathcal G_1(k_3,k_2;k_1)]
        \Big\}\ ,
        \eqa
where \bqa
        g_1 & \equiv & \frac{4 k_3 \prod_i k_i^2}
                            {(k_1 + k_2 - k_3)^2 k_t^2}\,, \nonumber \\
        g_2 & \equiv & \frac{4 k_2\prod_i k_i^2}
                            {(k_2^2 - (k_1 - k_3)^2)^2}\,, \nonumber \\
        g_3 & \equiv & \frac{4 k_1 \prod_i k_i^2}
                            {(k_1^2 - (k_2 + k_3)^2)^2}\,.
    \eqa
Then we find the summation of the three-point correlator. For
simplicity, the $V_{,\,IJK}$ terms are neglected.
    \bqa
\langle \delta\phi^{I}(\textbf{k}_1) \delta\phi^J(\textbf{k}_2)
        \delta\phi^K(\textbf{k}_3) \rangle & \supseteq &
        (2\pi)^3 \delta(\textbf{k}_1 + \textbf{k}_2 + \textbf{k}_3) \times \mbox{} \nonumber\\
        &&\frac{2 \pi^4 \mathrm P_\ast^2}{ \prod_i k_i^3}
        \sum_{perms}\frac{\dot{\phi}^I_\ast}{H_\ast}\delta^{JK}
        \mathcal A(k_1,k_2,k_3) ,
    \eqa
    \be \mathcal A(k_1,k_2,k_3)=  \frac{1}{2} k_1^3 -\frac{k_1(k_2^2+k_3^2)}{2}
           - \frac{4}{k_t} k_2^2 k_3^2 \ , \ee
where $\mathrm P_\ast =H_\ast^2/4\pi^2$ is the power spectrum of
scalar field when modes of scalar field perturbation cross the
horizon, and $\mathcal A(k_1,k_2,k_3)$ is the shape factor of the
three-point correlator. The result is easily reduced to the case of
single field, and consistent with~\cite{NG,eom}. \footnote{Since we
have used the metric of field space $\delta^{JK}$, the result is a
little different from~\cite{multi ng}. Therefore, $k_2$ and $k_3$
are symmetric here. If we symmetrize $k_2$ and $k_3$ in eq.(69)
of~\cite{multi ng}, then we get the same result. So both results are
equivalent.}

\section{Non-Gaussianity $f_{\rm NL}$}
\label{ng}

The non-Gaussianity of multiple-field inflation is shown in this
section. The inflaton will decay at the end of inflation, and the
final observable is the curvature perturbation $\zeta$. To discuss
the non-Gaussianity, the power spectrum and bispectrum of curvature
are given, \bqa \langle \zeta(\textbf{k}_1) \zeta(\textbf{k}_2)
\rangle &\equiv& (2 \pi)^3
\delta(\textbf{k}_1+\textbf{k}_2)\frac{2\pi^2}{k_1^3}\mathrm
P_{\zeta}(k_1)\ , \\
\langle \zeta(\textbf{k}_1) \zeta(\textbf{k}_2) \zeta(\textbf{k}_1)
\rangle &\equiv& (2\pi)^3
\delta(\textbf{k}_1+\textbf{k}_2+\textbf{k}_3)B_\zeta(k_1,k_2,k_3)\
. \eqa The non-Gaussianity is the deviation from the Gaussian
statistics in CMB, and use the parameter $f_{\rm NL}$ to represent
its magnitude, \be \zeta=\zeta_g+\frac{3}{5}f_{\rm
NL}\left({\zeta}_g^2-\langle {\zeta}_g^2\rangle\right) ~, \ee where
$\zeta_g$ denotes the Gaussian part of $\zeta$. \footnote{Here the
sign of $f_{\rm NL}$ is consistent with the convention of the CMB
experiments, and different from the paper written by
Maldacena~\cite{NG}. 
} Using this definition, we obtain \be \label{f}\frac{6}{5}f_{\rm
NL}=\frac{\prod_i k_i^3}{\sum_ik_i^3}\frac{B_\zeta}{4\pi^4\mathrm
P_\zeta^2} \ .\ee

Notice that the field equation formalism which is used to calculate
the non-Gaussianity is applicable when the modes cross the horizon.
In multiple-field inflation, there exists entropy perturbation. Thus
in order to consider the effects afterward, the $\delta N$ formalism
is a good method. On large scale, the value of curvature
perturbation is the  e-folding number from the initial flat slice at
$t_\ast$ to the final uniform density slice at time $t$, \be
\zeta(t,\textbf{x}) \simeq \delta N= N(t,
t_\ast,\textbf{x})-N(t,t_\ast)\ ,\ee where the e-folding number is
defined as \be N(t, t_\ast) \equiv \int^{t}_{t_\ast} H \d t \ .\ee
$\delta N$ can be expanded by the initial scalar fields, \be \delta
N= N_{,\,I}\delta \phi^{I}+\frac{1}{2} N_{,\,IJ} \delta
\phi^{I}\delta \phi^{J}+ \cdots\ . \ee The power spectrum and
bispectrum can be expressed by $\delta N$ formalism, \bqa \mathrm
P_{\zeta}&=&\delta^{IJ}N_{,\,I}N_{,\,J} \mathrm P_\ast \ , \label{1}
\\
\langle \zeta(\textbf{k}_1) \zeta(\textbf{k}_2) \zeta(\textbf{k}_3)
\rangle &=& N_{,\,I}N_{,\,J}N_{,\,K} \langle
\delta\phi^{I}(\textbf{k}_1) \delta\phi^J(\textbf{k}_2)
        \delta\phi^K(\textbf{k}_3) \rangle+ \nonumber\\
       && \hspace{-2cm}\frac{1}{2}N_{,\,I}N_{,\,J}N_{,\,KL}  \langle \delta\phi^{I}(\textbf{k}_1)
\delta\phi^J(\textbf{k}_2)
        (\delta\phi^K * \delta\phi^L)(\textbf{k}_3) \rangle +perms
        \ ,\label{2}
\eqa
where * denotes a convolution and the higher order terms are
neglected. The non-linear parameter $f_{\rm NL}$ is derived from
(\ref{f}), (\ref{1}) and (\ref{2}), \bqa f_{\rm NL} = \frac{5\rm
P_\ast}{12 \rm P_\zeta}\frac{1}{\sum_i k_i^3}(-\frac{1}{2} \sum_i
k_i^3 + \frac{4}{k_t} \sum_{i < j} k_i^2 k_j^2 + \frac{1}{2} \sum_{i
\neq j} k_i k_j^2)
+\frac{5}{6}\frac{N_{,\,I}N_{,\,J}N_{,\,IJ}}{(\delta^{IJ}N_{,\,I}N_{,\,J})^2}\
. \eqa The last term on the right hand side is from the curvature
evolution on large scale, which contributes the local form of
non-Gaussianity. The equilateral shape of non-Gaussianity from
multiple-field is constrained by the tensor-to-scalar ratio $r \sim
\rm P_\ast/\rm P_\zeta$.

\section{Conclusion}

In this paper, we derive the second-order field equation of
multiple-field, and calculate the shape of non-Gaussianity for
multiple-field inflation with the method of the field equation. The
shape of Non-Gaussiantiy derives from the three-point correlator,
which implies the microphysics in the period of inflation. Our
result of the three-point correlator is consistent with the previous
one~\cite{multi ng} which uses the method of in-in formalism. And it
is easy to extend the Bunch-Davies vacuum to the $\alpha$ vacuum in
the field equation formalism which shows the trans-Plankian
physics~\cite{trans} from non-Gaussianity.

The field equation formalism is applicable when we know the equation
of motion, even if the action is not given. Meanwhile, we should
notice that the formalism is used when the modes of scalar field
perturbation crossing the horizon. After crossing the horizon, the
quantum fluctuations become classical due to
decoherence~\cite{David}. Then the classical evolution of curvature
perturbation on large scale could be solved by the $\delta N$
formalism. Finally, we could get the non-Gaussianity observed in the
CMB.

\section*{Acknowledgments}
We are grateful to Yifu Cai for reading and revising the draft, and
thank Bin Chen, David Seery, Bo-Qiang Ma, Zhibo Xu for discussions
and communication.

\end{document}